\documentclass[10pt,conference]{IEEEtran}
\IEEEoverridecommandlockouts
\usepackage{cite}
\usepackage{multicol}
\usepackage{bm}
\usepackage{float}
\usepackage{array}
\usepackage{multirow}
\usepackage[cmex10]{amsmath}
\usepackage{graphicx}
\usepackage{url}
\usepackage{color}
\usepackage{lipsum}
\usepackage{cuted}
\usepackage{amssymb}
\usepackage{amsfonts}
\usepackage{amsmath}
\usepackage{extarrows}
\usepackage{graphicx}
\usepackage{amsfonts}
\usepackage{caption}
\usepackage{algorithm}
\usepackage{algpseudocode}
\usepackage{indentfirst}
\usepackage{caption}
\usepackage{esint} 
\usepackage{graphicx}
\usepackage{graphics}
\usepackage{pgfplots}
\usepackage{tikz}
\usepackage{epsfig}
\usepackage{soul}
\usepackage{caption}
\usepackage{gensymb}
\usepackage{geometry}
\def\BibTeX{{\rm B\kern-.05em{\sc i\kern-.025em b}\kern-.08em T\kern-.1667em\lower.7ex\hbox{E}\kern-.125emX}}

\DeclareMathOperator*{\argmax}{argmax}
\def\bA{{\mathbf{A}}} \def\bB{{\mathbf{B}}} \def\bC{{\mathbf{C}}} \def\bD{{\mathbf{D}}} 
\def\bF{{\mathbf{F}}}  \def\bH{{\mathbf{H}}} \def\bI{{\mathbf{I}}} \def\bJ{{\mathbf{J}}}
   \def\bN{{\mathbf{N}}} 
    \def\bT{{\mathbf{T}}}
    \def\bY{{\mathbf{Y}}}

\def\ba{{\mathbf{a}}}    
\def\bf{{\mathbf{f}}} \def\bg{{\mathbf{g}}} \def\bh{{\mathbf{h}}}  
   \def\bn{{\mathbf{n}}} 
\def\bp{{\mathbf{p}}}   \def\bs{{\mathbf{s}}} 
   \def\bx{{\mathbf{x}}} \def\by{{\mathbf{y}}}

\begin{document}
\title{Device-Free 3D Drone Localization in RIS-Assisted mmWave MIMO Networks}

\author{\IEEEauthorblockN{Jiguang~He$^{1,2}$, Charles Vanwynsberghe$^2$, Hui Chen$^{3}$, Chongwen Huang$^4$, and Aymen~Fakhreddine$^5$}

\IEEEauthorblockA{$^1$School of Computing and Information Technology, Great Bay University, Dongguan, China}

\IEEEauthorblockA{$^2$Technology Innovation Institute, 9639 Masdar City, Abu Dhabi, United Arab Emirates}

\IEEEauthorblockA{$^3$Department of Electrical Engineering, Chalmers University of Technology, Gothenburg, Sweden}

\IEEEauthorblockA{$^4$College of Information Science and Electronic
Engineering, Zhejiang University, Hangzhou 310027, China}

\IEEEauthorblockA{$^5$Institute of Networked and Embedded Systems, University of Klagenfurt, Klagenfurt, Austria}

 }

 \maketitle
\begin{abstract}
In this paper, we investigate the potential of reconfigurable intelligent surfaces (RISs) in facilitating passive/device-free three-dimensional (3D) drone localization within existing cellular infrastructure operating at millimeter-wave (mmWave) frequencies and employing multiple antennas at the transceivers. The developed localization system operates in the bi-static mode without requiring direct
communication between the drone and the base station. We analyze the theoretical performance limits via Fisher information analysis and Cram\'er Rao lower bounds (CRLBs). Furthermore, we develop a low-complexity yet effective drone localization algorithm based on coordinate gradient descent and examine the impact of factors such as radar cross section (RCS) of the drone and training overhead on system performance. It is demonstrated that integrating RIS yields significant benefits over its RIS-free counterpart, as evidenced by both theoretical analyses and numerical simulations.  
\end{abstract}
\section{Introduction}
Radio localization presents various formats depending on the interaction between the target and the base station (BS). When the target actively communicates with the BS, it constitutes active localization, commonly utilized in fifth generation (5G) and ultra-wideband (UWB) localization~\cite{Abu-Shaban2018}. Conversely, passive localization approaches, prevalent in radar sensing applications, come into play when the target does not communicate with the radar or BS. In addition, localization systems can be further categorized into mono-static or bi-static mode based on the collocation of the emitter/transmitter and receiver of pilot signals.

In the existing cellular systems, radar sensing is gaining tremendous momentum owing to the emergence of integrated communication and sensing (ISAC) paradigms~\cite{Liu2022,hua2024}. This innovation enables cellular BSs, traditionally focusing on communications, to incorporate sensing functionalities such as drone detection, localization, and tracking~\cite{Solomitckii2018}. In our latest work~\cite{he2024RIS}, we investigated drone detection leveraging reconfigurable intelligent surfaces (RISs) in millimeter-wave (mmWave) multiple-input multiple-output (MIMO) networks and examined the impact of various system parameters, including radar cross section (RCS)~\cite{Semkin2020}, on detection probability. Additional works in this domain can be found in~\cite{Buzzi2021,Buzzi2022}.

In this paper, we further extend our previous work to device-free three-dimensional (3D) drone localization, leveraging an RIS to enhance its performance within a bi-static setup. We develop a practical localization algorithm with low computational complexity based on coordinate gradient descent (CGD) while achieving performance close to the theoretical limits characterized by Fisher information analysis and the Cram\'er Rao lower bound (CRLB). Simulation results verify the feasibility of passive drone localization within cellular mmWave MIMO networks assisted by RISs, showcasing the advantages brought through the introduction of RISs.

\section{System Model}
\begin{figure}[t]
	\centering
\includegraphics[width=0.95\linewidth]{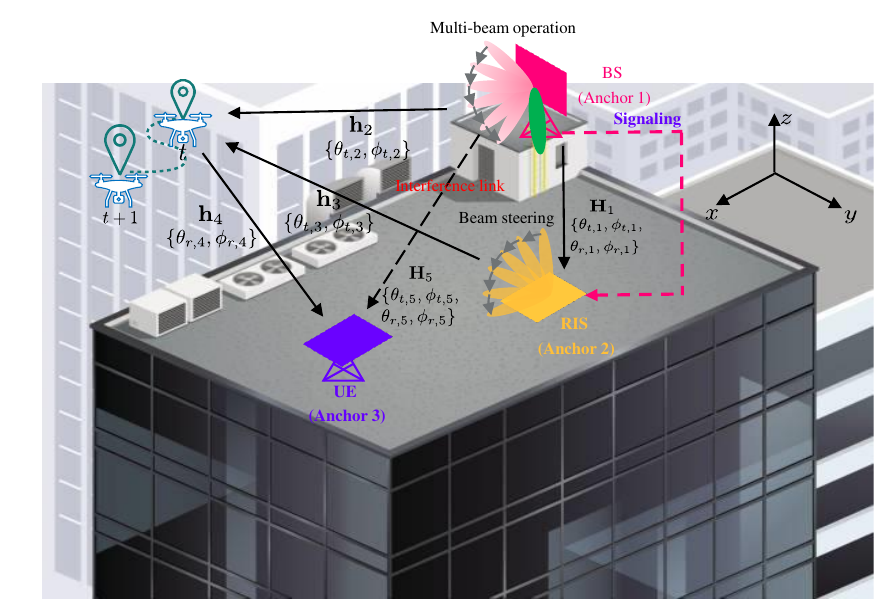}
	\caption{Device-free drone localization system consists of a mmWave BS, an RIS, and a receiving UE, deployed on the roof of a high-rise building.}
		\label{System_Model}
\end{figure}

Fig.~\ref{System_Model} showcases a novel device-free drone localization system deployed on the roof of a high-rise building. It comprises a mmWave BS, a passive RIS without any radio frequency (RF) chains and baseband processing units, and a receiving user equipment (UE), all possessing a uniform planar array (UPA) structure with multiple antennas or elements. This configuration resembles a conventional bi-static radar system with spatial separation of the transmitter and receiver, but harnesses augmented capabilities through the introduction of an RIS.  All the three network nodes serve as positioning reference nodes, a.k.a. anchors. The BS array aligns with the $y\text{-}z$ plane, while the RIS and UE arrays are oriented parallel to the $x\text{-}y$ plane, facing towards the sky.\footnote{\textit{Notations}: A bold lowercase letter $\ba$ denotes a vector, and a bold capital letter $\bA$ denotes a matrix. The operators $(\cdot)^*$, $(\cdot)^\mathsf{T}$, $(\cdot)^\mathsf{H}$, and $(\cdot)^{-1}$ denote the matrix or vector conjugate, transpose, Hermitian transpose, and inverse, respectively. $[\ba]_m$, $[\bA]_{mn}$, and $[\bA]_{:, m:n}$ denote the $m$th element of $\ba$, the $(m,n)$th element of $\bA$, and the submatrix of $\bA$ formed by its elements in the columns $m,m+1, \ldots, n$. $\mathrm{diag}(\ba)$ denotes a square diagonal matrix with the entries of $\ba$ on its diagonal, $\mathrm{vec}(\bA)$ denotes the vectorization of $\bA$ by stacking the columns of the matrix $\bA$ on top of one another,
$\bA \otimes \bB$ denotes the Kronecker product of $\bA$ and $\bB$, $\mathrm{rank} (\bA)$ returns the rank of matrix $\bA$, $\mathrm{Tr}(\bA)$ denotes the trace of $\bA$, $\mathbb{E}[\cdot]$ is the expectation operator, $\mathbf{0}$ denotes the all-zero vector, $\bI_{M}$ denotes the $M\times M$ identity matrix, and $j = \sqrt{-1}$. $\|\cdot\|_\mathrm{F}$ denotes the Frobenius norm of a matrix, $|\cdot|$ returns the absolute value of a complex number, and $\|\cdot\|_2$ denotes the Euclidean norm of a vector. Finally, $\Re\{\cdot\}$ returns the real part of its complex argument.} To improve sounding efficiency, the BS employs a multi-beam approach with one fixed beam towards the RIS and one time-varying beam towards the sky. The primary goal of such a system is to localize a flying drone with a relatively low RCS, without requiring information exchange between the drone and the BS. 
\subsection{mmWave Channel Model}
In contrast to sub-6GHz wireless channels, mmWave channels exhibit poor-scattering property. Furthermore, since we consider an open space scenario, we focus solely on the line-of-sight (LoS) path when modelling the mmWave channels. Intuitively, the RIS should be placed in the proximity of the BS, with the potential to reduce the path loss. The channel between the BS and the RIS, denoted as $\bH_1\in\mathbb{C}^{M_\text{R}\times M_\text{B}}$, where $M_\text{R}$ and $M_\text{B}$ represent the number of RIS elements and BS antennas respectively, can be effectively characterized using the Saleh-Valenzuela channel model~\cite{Heath2016}, a widely adopted approach, as 
 \begin{align}\label{H_1}
\bH_1 =& \frac{e^{-j 2\pi  d_1/\lambda}}{\sqrt{\rho_1}} \boldsymbol{\alpha}_x(\theta_{r,1},\phi_{r,1}) \otimes \boldsymbol{\alpha}_y(\theta_{r,1},\phi_{r,1})\nonumber\\ &\big(\boldsymbol{\alpha}_y(\theta_{t,1},\phi_{t,1}) \otimes \boldsymbol{\alpha}_z(\phi_{t,1})\big)^\mathsf{H},
\end{align}
 where $\rho_1$ represents the path loss, and $d_1$ denotes the distance between the BS and RIS. The term $\lambda$ corresponds to the wavelength of the carrier frequency, while $\theta_{r,1}$ ($=\theta_{t,1}$) and $\phi_{r,1}$ ($=\phi_{t,1}$) are the azimuth and elevation angles of arrival (departure) associated with $\bH_1$, respectively. Here, the subscripts $t$ and $r$ denote the transmitter and receiver, respectively. To be specific, the array response vectors $\boldsymbol{\alpha}_x(\theta_{r,1},\phi_{r,1})$, $\boldsymbol{\alpha}_y(\theta_{r,1},\phi_{r,1})$, $\boldsymbol{\alpha}_y(\theta_{t,1},\phi_{t,1})$,  and $\boldsymbol{\alpha}_z(\phi_{t,1})$ can be written as~\cite{Tsai2018}
 \begin{align}
    & \boldsymbol{\alpha}_x(\theta_{r,1},\phi_{r,1}) = \Big[e^{-j \frac{2\pi  d_{\text{R},x}}{\lambda} \big(\frac{M_{\text{R},x} -1}{2}\big) \cos(\theta_{r,1}) \sin(\phi_{r,1})}, \nonumber\\
    & \cdots, e^{j \frac{2\pi d_{\text{R},x}}{\lambda} \big(\frac{M_{\text{R},x} -1}{2}\big) \cos(\theta_{r,1})\sin(\phi_{r,1})} \Big]^{\mathsf{T}},
    \end{align}
  
    \begin{align}    &\boldsymbol{\alpha}_y(\theta_{r,1},\phi_{r,1}) = \Big[e^{-j \frac{2\pi  d_{{\text{R},y}}}{\lambda} \big(\frac{M_{\text{R},y} -1}{2}\big) \sin(\theta_{r,1}) \sin(\phi_{r,1})}, \nonumber\\
    & \cdots, e^{j \frac{2\pi d_{{\text{R},y}}}{\lambda} \big(\frac{M_{{\text{R},y}} -1}{2}\big) \sin(\theta_{r,1})\sin(\phi_{r,1})} \Big]^{\mathsf{T}}, \\ &\boldsymbol{\alpha}_y(\theta_{t,1},\phi_{t,1}) = \Big[e^{-j \frac{2\pi  d_{{\text{B},y}}}{\lambda} \big(\frac{M_{{\text{B},y}} -1}{2}\big) \sin(\theta_{t,1}) \sin(\phi_{t,1})}, \nonumber\\
    & \cdots, e^{j \frac{2\pi d_{{\text{B},y}}}{\lambda} \big(\frac{M_{{\text{B},y}} -1}{2}\big) \sin(\theta_{t,1})\sin(\phi_{t,1})} \Big]^{\mathsf{T}},\\
       &\boldsymbol{\alpha}_z(\phi_{t,1}) = \Big[e^{-j \frac{2\pi d_{\text{B},z}}{\lambda} \big(\frac{M_{\text{B},z} -1}{2}\big) \cos(\phi_{t,1}) }, \nonumber\\
    & \cdots, e^{j \frac{2\pi d_{\text{B},z}}{\lambda} \big(\frac{M_{\text{B},z} -1}{2}\big) \cos(\phi_{t,1})} \Big]^{\mathsf{T}}, 
 \end{align}  
 where $M_\text{R} = M_{\text{R},x} M_{\text{R},y}$ with $M_{\text{R},x}$ and $M_{\text{R},y}$ being the number of RIS elements along the $x$ axis and $y$ axis, respectively. Additionally, $d_{\text{R},x}$ and $d_{\text{R},y}$ represent the inter-element spacing for RIS elements along the $x$-axis and $y$-axis respectively. Similarly, $M_\text{B} = M_{\text{B},y} \times M_{\text{B},z}$, with $M_{\text{B},y}$ and $M_{\text{B},z}$ denoting the number of BS antennas along the $y$-axis and $z$-axis, respectively. Furthermore, $d_{\text{B},y}$ and $d_{\text{B},z}$ represent the inter-element spacing for BS antennas along the $y$-axis and $z$-axis, respectively. 
 
 Similarly, the remaining channels (marked in Fig.~\ref{System_Model}), e.g., $\bh_2\in\mathbb{C}^{1\times M_\text{B}}$, $\bh_3\in\mathbb{C}^{1\times M_\text{R}}$, $\bh_4 \in\mathbb{C}^{M_\text{U}\times 1}$, and $\bH_5\in\mathbb{C}^{M_\text{U}\times M_\text{B}}$ can be presented in the same manner, listed below:
 \begin{align}
     \bh_2 &=\frac{e^{-j 2\pi  d_2/\lambda}}{\sqrt{\rho_2}} \big(\boldsymbol{\alpha}_y(\theta_{t,2},\phi_{t,2}) \otimes \boldsymbol{\alpha}_z(\phi_{t,2})\big)^\mathsf{H},\\
     \bh_3 &=\frac{e^{- j 2\pi   d_3/\lambda}}{\sqrt{\rho_3}} \big(\boldsymbol{\alpha}_x(\theta_{t,3},\phi_{t,3}) \otimes \boldsymbol{\alpha}_y(\theta_{t,3},\phi_{t,3})\big)^\mathsf{H},\\
     \bh_4 &=\frac{e^{-j 2\pi  d_4/\lambda}}{\sqrt{\rho_4}} \boldsymbol{\alpha}_x(\theta_{r,4},\phi_{r,4}) \otimes \boldsymbol{\alpha}_y(\theta_{r,4},\phi_{r,4}), \\
     \bH_5 &= \frac{e^{-j 2\pi  d_5/\lambda}}{\sqrt{\rho_5}} \boldsymbol{\alpha}_x(\theta_{r,5},\phi_{r,5}) \otimes \boldsymbol{\alpha}_y(\theta_{r,5},\phi_{r,5})\nonumber\\ &\hspace{0.47cm}\big(\boldsymbol{\alpha}_y(\theta_{t,5},\phi_{t,5}) \otimes \boldsymbol{\alpha}_z(\phi_{t,5})\big)^\mathsf{H},
 \end{align}
where $M_\text{U}$ is the number of UE antennas, $d_2(\rho_2)$, $d_3(\rho_3)$, and $d_4(\rho_4)$ are the distance (path loss) between the drone and the BS, the RIS, and the UE, respectively, $d_5(\rho_5)$ is the distance (path loss) between the BS and the UE. $\theta_{t,i}$ and $\phi_{t,i}$ are the azimuth and elevation angles of departure associated with $\bh_i$ for $i = 2,3$, and $\theta_{r,4}$ ($\theta_{r,5}$) and $\phi_{r,4}$ ($\phi_{r,5}$) are the azimuth and elevation angles of arrival associated with $\bh_4$~($\bH_5$). $\theta_{t,5}$ and $\phi_{t,5}$ are the azimuth and elevation angles of departure associated with $\bH_5$. In this device-free drone localization system, $\bH_5$ is the direct channel between the BS and the UE, which serves as an interference link. 
 
 \subsection{Downlink Sounding Procedure} \label{sec_sounding_procedure}
\begin{figure}[t]
	\centering
\includegraphics[width=0.9\linewidth]{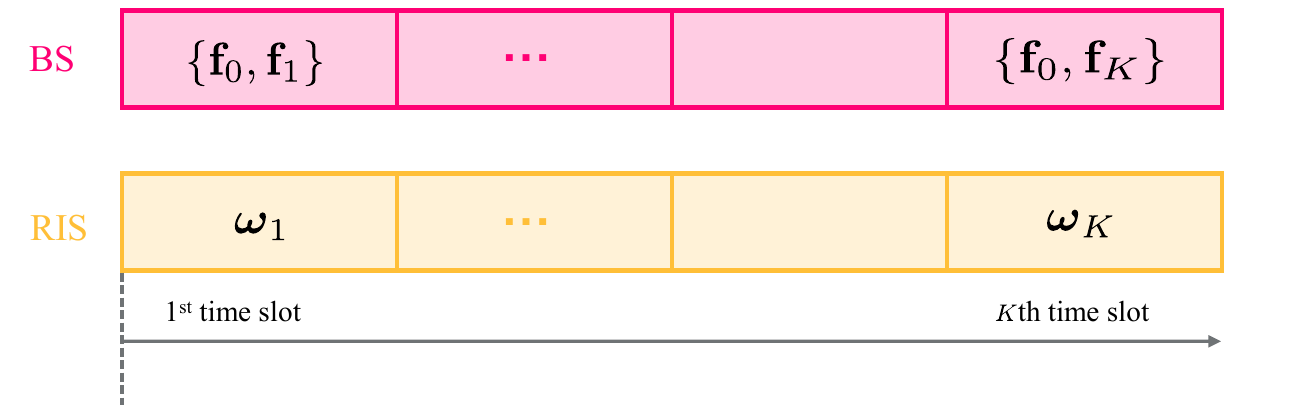}
	\caption{The sounding framework for passive drone localization, where two beams $\bf_0, \bf_k \in \mathbb{C}^{M_\text{B} \times 1}$ are concurrently employed at the BS, the former fixed beam towards the RIS and the latter time-varying beam towards the sky, and one time-varying beam $ \boldsymbol{\omega}_k \in \mathbb{C}^{M_\text{R} \times 1}$ is considered at the RIS towards the sky, during time slot $k$.}	\label{Sounding_Frame}
\end{figure}

The sounding procedure, illustrated in Fig.~\ref{Sounding_Frame}, involves the BS transmitting beamformed pilot signals over the downlink to illuminate both the RIS and a specific portion of the sky simultaneously, employing a multi-beam approach. Meanwhile, the RIS adjusts its beams to cover a designated area of the sky where the drone might be located. The BS beam $\mathbf{f}_0 \in \mathbb{C}^{M_\text{B}}$ ($\|\bf_0\|_2 = \sqrt{M_\text{B}}$) directed towards the RIS remains fixed over time, while the BS beam $\mathbf{f}_k\in \mathbb{C}^{M_\text{B}}$ ($\|\bf_k\|_2 = \sqrt{M_\text{B}}$) and the RIS beam $\boldsymbol{\omega}_k\in \mathbb{C}^{M_\text{R}}$ ($\|\boldsymbol{\omega}_k\|_2 = \sqrt{M_\text{R}}$) directed towards the sky vary with time, for $k = 1,\cdots, K$, with $K$ indicating the level of training overhead. We guarantee that the pairs of BS beams are mutually orthogonal, i.e., $\bf_0^\mathsf{H} \bf_k = 0 $, $\forall k$. Each element of RIS phase profile $\boldsymbol{\omega}_k$ satisfies unit-modulus constraint, i.e., $|[\boldsymbol{\omega}_k]_i| = 1$, for $\forall i$. The pilot signals transmitted from the BS traverse various paths to reach the UE, including a direct path (BS-UE), a single-bounce path (BS-drone-UE), and a double-bounce path (BS-RIS-drone-UE).


We assume that the exact locations of the RIS and UE are known to the BS. In this sense, the BS can take full advantage of such prior information and use it for the purpose of beam design during the sounding procedure.  
It should be noted that during the sounding procedure, the beam design at the BS and RIS is essential for the passive 3D drone localization systems. By following the maximum ratio transmission (MRT) principle based on the known positions of the BS and RIS, we set $\bf_0 = \big(\boldsymbol{\alpha}_y(\theta_{t,1},\phi_{t,1}) \otimes \boldsymbol{\alpha}_z(\phi_{t,1})\big)$. To further make the BS beams $\bf_0, \bf_k$ orthogonal and mitigate the interference from $\bH_5$, we define $\bg_0 =  \big(\boldsymbol{\alpha}_y(\theta_{t,5},\phi_{t,5}) \otimes \boldsymbol{\alpha}_z(\phi_{t,5})\big)$ and constrain the remaining beams (i.e., $\bf_k$, for $k =1,\cdots, K$) to lie within the null space of the matrix $\bF^\mathsf{H}$, where $\bF = [\bf_0 ,\; \bg_0]$. As a result, $\bf_k^\mathsf{H} \bf_0  = \bf_k^\mathsf{H} \bg_0 =0$, $\forall k$.

\subsection{Signal Model}
The received signal at the receiving UE during the time slot $k$ can be expressed as 
\begin{equation}\label{by_k}
    \by_k =  \zeta\bh_4 (\bh_3 \mathrm{diag}(\boldsymbol{\omega}_k) \bH_1 \bF_k \bs_k + \bh_2 \bF_k \bs_k) + \bH_5 \bF_k \bs_k +\bn_k,
\end{equation}
where $\zeta$ is the reflection coefficient of the drone, a.k.a. RCS, $\bF_k = [\bf_0,\, \bf_k]$, $\bs_k = [s_0,\, s_k]^\mathsf{T}$, and the transmitted pilot symbols during each time slot satisfy the following sum power constraint: $\mathbb{E}\{|s_0|^2\} + \mathbb{E}\{|s_k|^2\} = P$ for $\forall k$. Without loss of generality, we set $s_0 = s_k = \sqrt{P/2}$. The additive noise vector at the UE follows complex Gaussian distribution, i.e., $\bn_k \sim \mathcal{CN}(\mathbf{0},\sigma^2 \bI_{M_\text{U}})$. Note that the interference term $\bH_5 \bF_k \bs_k$ is equal to $\bH_5 \bf_0 s_0$ due to $\bH_5 \bf_k = \mathbf{0}$. Namely, the interference comes only from the beam $\bf_0$. By carefully choosing the locations of the RIS and UE, we can further reduce and even eliminate the interference term in Eq.~\eqref{by_k}. 
The signal vector $\by_k$ in~\eqref{by_k} can be further expressed as 
\begin{align}\label{by_k1}
    \by_k &=  \zeta\bh_4 (\bh_3\mathrm{diag}(\boldsymbol{\alpha}_x(\theta_{r,1},\phi_{r,1}) \otimes \boldsymbol{\alpha}_y(\theta_{r,1},\phi_{r,1})) \boldsymbol{\omega}_k \nonumber\\ &\times \frac{e^{-j 2\pi  d_1/\lambda}}{\sqrt{\rho_1}} \big(\boldsymbol{\alpha}_y(\theta_{t,1},\phi_{t,1}) \otimes \boldsymbol{\alpha}_z(\phi_{t,1})\big)^\mathsf{H} \bx_k \nonumber\\
    &+ \bh_2 \bx_k) + \bH_5 \bx_k + \bn_k \nonumber\\
    &=\sqrt{P/2} \frac{e^{-j 2\pi  d_1/\lambda}}{\sqrt{\rho_1}} M_\text{B} \zeta\bh_4 \bh_3 \nonumber\\ &\times \mathrm{diag}(\boldsymbol{\alpha}_x(\theta_{r,1},\phi_{r,1}) \otimes \boldsymbol{\alpha}_y(\theta_{r,1},\phi_{r,1}))\boldsymbol{\omega}_k \nonumber\\
    &+ \sqrt{P/2}\zeta\bh_4\bh_2 \bar{\bf}_k+ \sqrt{P/2}\bH_5 \bar{\bf}_k + \bn_k,
\end{align}
where $\bx_k = \bF_k \bs_k$ and $\bar{\bf}_k = \bf_0 + \bf_k$. 
By stacking $\{\by_k\}$ column by column, the received signal can be expressed as 
\begin{equation}\label{Rec_signal}
    \bY = \sqrt{P/2}\tilde{\bH} \bar{\boldsymbol{\Omega}} + \sqrt{P/2}\hat{\bH} \bar{\bF} + \sqrt{P/2}\bH_5\bar{\bF} +  \bN,
\end{equation}
where $\bar{\boldsymbol{\Omega}} = [ \mathrm{diag}(\boldsymbol{\alpha}_x(\theta_{r,1},\phi_{r,1}) \otimes \boldsymbol{\alpha}_y(\theta_{r,1},\phi_{r,1}))\boldsymbol{\omega}_1,  \cdots, \mathrm{diag}(\boldsymbol{\alpha}_x(\theta_{r,1},\phi_{r,1}) \otimes \boldsymbol{\alpha}_y(\theta_{r,1},\phi_{r,1}))\boldsymbol{\omega}_K ], \bar{\bF} = [ \bar{\bf}_1, \cdots,  \bar{\bf}_K]$, and $\bN = [\bn_1, \cdots, \bn_K]$. The two cascaded channels $ \tilde{\bH}$ and $ \hat{\bH}$ are function of $\zeta$, $\bh_4$, $\bh_3$, and $\bh_2$, as
\begin{align}\label{H_tilde}
   \tilde{\bH} &=  \zeta  M_\text{B} \frac{e^{-j 2\pi  d_1/\lambda}}{\sqrt{\rho_1}} \bh_4 \bh_3 
     = \tilde{\epsilon} \boldsymbol{\alpha}_x(\theta_{r,4},\phi_{r,4}) \boldsymbol{\alpha}_x(\theta_{t,3},\phi_{t,3})^\mathsf{H}\nonumber\\
     &\otimes \boldsymbol{\alpha}_y(\theta_{r,4},\phi_{r,4}) \boldsymbol{\alpha}_y(\theta_{t,3},\phi_{t,3})^\mathsf{H}, \\
   \hat{\bH} &=\zeta\bh_4\bh_2  = \hat{\epsilon} \boldsymbol{\alpha}_x(\theta_{r,4},\phi_{r,4}) \boldsymbol{\alpha}_y(\theta_{t,2},\phi_{t,2})^\mathsf{H} \nonumber\\
   &\otimes \boldsymbol{\alpha}_y(\theta_{r,4},\phi_{r,4}) \boldsymbol{\alpha}_z(\phi_{t,2})^\mathsf{H}, \label{H_hat}
\end{align}
where $\tilde{\epsilon} = \zeta  M_\text{B} \frac{e^{-j 2\pi  d_1/\lambda}}{\sqrt{\rho_1}} \frac{e^{-j2  \pi d_4/\lambda}}{\sqrt{\rho_4}} \frac{e^{-j2\pi d_3/\lambda}}{\sqrt{\rho_3}}$ and $\hat{\epsilon} =\zeta \frac{e^{-j2\pi d_4/\lambda}}{\sqrt{\rho_4}} \frac{e^{-j2\pi d_2/\lambda}}{\sqrt{\rho_2}}$.

Since we assume that the angular parameters in $\bH_1$ are known \textit{a priori} due to the known positions of the BS and RIS, $\mathrm{diag}(\boldsymbol{\alpha}_x(\theta_{r,1},\phi_{r,1}) \otimes \boldsymbol{\alpha}_y(\theta_{r,1},\phi_{r,1}))$ is known in $\bar{\boldsymbol{\Omega}}$. Therefore, in Eq.~\eqref{Rec_signal}, both $\bar{\boldsymbol{\Omega}}$ and $\bar{\bF}$ are known to the UE. Recall that the goal of the study is to localize the drone in a passive device-free manner based on the received signal $\bY$ at the UE. The unknown parameters in $\tilde{\bH}$ are $\bh_3$, $\bh_4$, $\zeta$, and $\frac{e^{- j 2\pi d_1/\lambda}}{\sqrt{\rho_1}}$. In~$\hat{\bH}$, all the three terms, i.e., $\zeta$, $\bh_4$, and $\bh_2$, are unknown. Both $\tilde{\bH}$ and $\hat{\bH}$ are rank-one matrices. In the considered localization system, we have three reference nodes, i.e., the BS, the UE, and the RIS. Provided that we can estimate the angular parameters in $\bh_2$, $\bh_3$, and $\bh_4$, i.e., $\theta_{t,2}$, $\phi_{t,2}$, $\theta_{t,3}$, $\phi_{t,3}$, $\theta_{r,4}$, and $\phi_{r,4}$, from the received signal $\bY$ in Eq.~\eqref{Rec_signal}, we can exploit the Triangulation localization method to calculate the 3D coordinate of the drone, realizing passive 3D drone localization. 

\section{Performance Limit Analysis}
In this section, we characterize the 3D drone localization via Fisher information analysis and CRLB. According to Eqs.~\eqref{H_tilde} and~\eqref{H_hat}, the unknown parameters in the received signal in~Eq.~\eqref{Rec_signal} are complex-valued $\hat{\epsilon}$ and $\tilde{\epsilon}$ as well as real-valued $\theta_{t,2}$, $\phi_{t,2}$, $\theta_{t,3}$, $\phi_{t,3}$, $\theta_{r,4}$, and $\phi_{r,4}$. We stack all the unknown parameters into a vector $\boldsymbol{\eta} = [\Re(\tilde{\epsilon}) \; \Im(\tilde{\epsilon}) \;\Re(\hat{\epsilon}) \;\Im(\hat{\epsilon}) \;  \theta_{t,2}\; \phi_{t,2}\;\theta_{t,3}\; \phi_{t,3}\;\theta_{r,4}\;  \phi_{r,4}]^\mathsf{T} \in \mathbb{R}^{10}$. The vectorized received signal can be expressed as 
\begin{align}\label{eq:rec_sig}
\by = &\sqrt{P/2} \tilde{\epsilon} (\bar{\boldsymbol{\Omega}}^\mathsf{T} \otimes \bI_{M_\text{U}} )(\boldsymbol{\alpha}_x^*(\theta_{t,3},\phi_{t,3}) \otimes \boldsymbol{\alpha}_y^*(\theta_{t,3},\phi_{t,3}) \nonumber\\
&\otimes \boldsymbol{\alpha}_x(\theta_{r,4},\phi_{r,4}) \otimes \boldsymbol{\alpha}_y(\theta_{r,4},\phi_{r,4}))\nonumber\\
&+ \sqrt{P/2} \hat{\epsilon} (\bar{\bF}^\mathsf{T} \otimes \bI_{M_\text{U}} )(\boldsymbol{\alpha}_y^*(\theta_{t,2},\phi_{t,2}) \otimes \boldsymbol{\alpha}_z^*(\phi_{t,2})\nonumber\\
&\otimes \boldsymbol{\alpha}_x(\theta_{r,4},\phi_{r,4}) \otimes \boldsymbol{\alpha}_y(\theta_{r,4},\phi_{r,4}))\hspace{-0.07cm}+\hspace{-0.07cm} \sqrt{\frac{P}{2}}\mathrm{vec}(\bH_5\bar{\bF})\hspace{-0.07cm} + \hspace{-0.07cm}\bn,
\end{align}
where $\by = \mathrm{vec}(\bY)$, and $\bn = \mathrm{vec}(\bN)$. The Fisher information matrix $\bJ \in \mathbb{C}^{10 \times 10}$ for $\boldsymbol{\eta}$ is written as~\cite{Jiguang2023}
\begin{equation}
[\bJ]_{mn} = \frac{2}{\sigma^2} \Re\Big\{ \frac{\partial \boldsymbol{\mu}^\mathsf{H}}{[\boldsymbol{\eta}]_m} \frac{\partial \boldsymbol{\mu}}{[\boldsymbol{\eta}]_n} \Big\},
\end{equation}
where $\boldsymbol{\mu}$ is the noise- and interference-free signal from Eq.~\eqref{eq:rec_sig}. The details on $\frac{\partial \boldsymbol{\mu}}{[\boldsymbol{\eta}]_n}$ are provided in Eqs.~\eqref{eq:partial_derivative_begin}--\eqref{eq:partial_derivative_end},
\begin{figure*}\small
\begin{align}\label{eq:partial_derivative_begin}
    \frac{\partial \boldsymbol{\mu}}{[\boldsymbol{\eta}]_1} &=\sqrt{P/2} (\bar{\boldsymbol{\Omega}}^\mathsf{T} \otimes \bI_{M_\text{U}} )(\boldsymbol{\alpha}_x^*(\theta_{t,3},\phi_{t,3}) \otimes \boldsymbol{\alpha}_y^*(\theta_{t,3},\phi_{t,3}) \otimes \boldsymbol{\alpha}_x(\theta_{r,4},\phi_{r,4}) \otimes \boldsymbol{\alpha}_y(\theta_{r,4},\phi_{r,4})), \\
    \frac{\partial \boldsymbol{\mu}}{[\boldsymbol{\eta}]_2} &=j \sqrt{P/2} (\bar{\boldsymbol{\Omega}}^\mathsf{T} \otimes \bI_{M_\text{U}} )(\boldsymbol{\alpha}_x^*(\theta_{t,3},\phi_{t,3}) \otimes \boldsymbol{\alpha}_y^*(\theta_{t,3},\phi_{t,3}) \otimes \boldsymbol{\alpha}_x(\theta_{r,4},\phi_{r,4}) \otimes \boldsymbol{\alpha}_y(\theta_{r,4},\phi_{r,4})),    \\
    \frac{\partial \boldsymbol{\mu}}{[\boldsymbol{\eta}]_3}&= \sqrt{P/2}  (\bar{\bF}^\mathsf{T} \otimes \bI_{M_\text{U}} )(\boldsymbol{\alpha}_y^*(\theta_{t,2},\phi_{t,2}) \otimes \boldsymbol{\alpha}_z^*(\phi_{t,2}) \otimes \boldsymbol{\alpha}_x(\theta_{r,4},\phi_{r,4}) \otimes \boldsymbol{\alpha}_y(\theta_{r,4},\phi_{r,4})),\\
    \frac{\partial \boldsymbol{\mu}}{[\boldsymbol{\eta}]_4}&= j \sqrt{P/2}  (\bar{\bF}^\mathsf{T} \otimes \bI_{M_\text{U}} )(\boldsymbol{\alpha}_y^*(\theta_{t,2},\phi_{t,2}) \otimes \boldsymbol{\alpha}_z^*(\phi_{t,2}) \otimes \boldsymbol{\alpha}_x(\theta_{r,4},\phi_{r,4}) \otimes \boldsymbol{\alpha}_y(\theta_{r,4},\phi_{r,4})),\\
    \frac{\partial \boldsymbol{\mu}}{[\boldsymbol{\eta}]_5}&=  \sqrt{P/2} \hat{\epsilon} (\bar{\bF}^\mathsf{T} \otimes \bI_{M_\text{U}} )(\boldsymbol{\Phi}_1 \boldsymbol{\alpha}_y^*(\theta_{t,2},\phi_{t,2}) \otimes \boldsymbol{\alpha}_z^*(\phi_{t,2}) \otimes \boldsymbol{\alpha}_x(\theta_{r,4},\phi_{r,4}) \otimes \boldsymbol{\alpha}_y(\theta_{r,4},\phi_{r,4})),\\
   \frac{\partial \boldsymbol{\mu}} {[\boldsymbol{\eta}]_6}&= \sqrt{P/2} \hat{\epsilon} (\bar{\bF}^\mathsf{T} \otimes \bI_{M_\text{U}} )((\boldsymbol{\Phi}_2 \boldsymbol{\alpha}_y^*(\theta_{t,2},\phi_{t,2}) \otimes \boldsymbol{\alpha}_z^*(\phi_{t,2}) +  \boldsymbol{\alpha}_y^*(\theta_{t,2},\phi_{t,2}) \otimes \boldsymbol{\Phi}_3\boldsymbol{\alpha}_z^*(\phi_{t,2}))\otimes \boldsymbol{\alpha}_x(\theta_{r,4},\phi_{r,4}) \otimes \boldsymbol{\alpha}_y(\theta_{r,4},\phi_{r,4})),\\
    \frac{\partial \boldsymbol{\mu}}{[\boldsymbol{\eta}]_7} \hspace{-0.1cm}&=\hspace{-0.1cm}\sqrt{\frac{P}{2}} \tilde{\epsilon}(\bar{\boldsymbol{\Omega}}^\mathsf{T} \otimes \bI_{M_\text{U}} )((\boldsymbol{\Phi}_4\boldsymbol{\alpha}_x^*(\theta_{t,3},\phi_{t,3}) \otimes \boldsymbol{\alpha}_y^*(\theta_{t,3},\phi_{t,3})\hspace{-0.1cm} +\hspace{-0.1cm}(\boldsymbol{\alpha}_x^*(\theta_{t,3},\phi_{t,3}) \otimes \boldsymbol{\Phi}_5\boldsymbol{\alpha}_y^*(\theta_{t,3},\phi_{t,3}))\otimes \boldsymbol{\alpha}_x(\theta_{r,4},\phi_{r,4}) \otimes \boldsymbol{\alpha}_y(\theta_{r,4},\phi_{r,4})), \\
    \frac{\partial \boldsymbol{\mu}}{[\boldsymbol{\eta}]_8}\hspace{-0.1cm} &=\hspace{-0.1cm}\sqrt{\frac{P}{2}} \tilde{\epsilon}(\bar{\boldsymbol{\Omega}}^\mathsf{T} \otimes \bI_{M_\text{U}} )((\boldsymbol{\Phi}_6\boldsymbol{\alpha}_x^*(\theta_{t,3},\phi_{t,3}) \otimes \boldsymbol{\alpha}_y^*(\theta_{t,3},\phi_{t,3}) \hspace{-0.1cm}+\hspace{-0.1cm}(\boldsymbol{\alpha}_x^*(\theta_{t,3},\phi_{t,3}) \otimes \boldsymbol{\Phi}_7\boldsymbol{\alpha}_y^*(\theta_{t,3},\phi_{t,3}))\otimes \boldsymbol{\alpha}_x(\theta_{r,4},\phi_{r,4}) \otimes \boldsymbol{\alpha}_y(\theta_{r,4},\phi_{r,4})), \\
    \frac{\partial \boldsymbol{\mu}}{[\boldsymbol{\eta}]_9} \hspace{-0.1cm}&=\hspace{-0.1cm}\sqrt{\frac{P}{2}} \tilde{\epsilon}(\bar{\boldsymbol{\Omega}}^\mathsf{T} \otimes \bI_{M_\text{U}} )(\boldsymbol{\alpha}_x^*(\theta_{t,3},\phi_{t,3}) \otimes \boldsymbol{\alpha}_y^*(\theta_{t,3},\phi_{t,3}) \otimes (\boldsymbol{\Phi}_8\boldsymbol{\alpha}_x(\theta_{r,4},\phi_{r,4}) \otimes \boldsymbol{\alpha}_y(\theta_{r,4},\phi_{r,4})\hspace{-0.1cm}+\hspace{-0.1cm}\boldsymbol{\alpha}_x(\theta_{r,4},\phi_{r,4}) \otimes \boldsymbol{\Phi}_9\boldsymbol{\alpha}_y(\theta_{r,4},\phi_{r,4}))) \nonumber\\
    &+\sqrt{\frac{P}{2}} \hat{\epsilon} (\bar{\bF}^\mathsf{T} \otimes \bI_{M_\text{U}} )(\boldsymbol{\alpha}_y^*(\theta_{t,2},\phi_{t,2}) \otimes \boldsymbol{\alpha}_z^*(\phi_{t,2}) \otimes (\boldsymbol{\Phi}_8\boldsymbol{\alpha}_x(\theta_{r,4},\phi_{r,4}) \otimes \boldsymbol{\alpha}_y(\theta_{r,4},\phi_{r,4})+\boldsymbol{\alpha}_x(\theta_{r,4},\phi_{r,4}) \otimes \boldsymbol{\Phi}_9\boldsymbol{\alpha}_y(\theta_{r,4},\phi_{r,4}))),\\
        \frac{\partial \boldsymbol{\mu}}{[\boldsymbol{\eta}]_{10}} \hspace{-0.1cm}&=\hspace{-0.1cm}\sqrt{\frac{P}{2}} \tilde{\epsilon}(\bar{\boldsymbol{\Omega}}^\mathsf{T} \hspace{-0.05cm}\otimes\hspace{-0.05cm} \bI_{M_\text{U}} )(\boldsymbol{\alpha}_x^*(\theta_{t,3},\phi_{t,3})\hspace{-0.05cm} \otimes \hspace{-0.05cm}\boldsymbol{\alpha}_y^*(\theta_{t,3},\phi_{t,3}) \hspace{-0.05cm}\otimes\hspace{-0.05cm} (\boldsymbol{\Phi}_{10}\boldsymbol{\alpha}_x(\theta_{r,4},\phi_{r,4}) \hspace{-0.05cm}\otimes\hspace{-0.05cm} \boldsymbol{\alpha}_y(\theta_{r,4},\phi_{r,4})\hspace{-0.1cm}+\hspace{-0.1cm}\boldsymbol{\alpha}_x(\theta_{r,4},\phi_{r,4}) \hspace{-0.05cm}\otimes\hspace{-0.05cm} \boldsymbol{\Phi}_{11}\boldsymbol{\alpha}_y(\theta_{r,4},\phi_{r,4}))) \nonumber\\
    &+\sqrt{\frac{P}{2}} \hat{\epsilon} (\bar{\bF}^\mathsf{T} \otimes \bI_{M_\text{U}})(\boldsymbol{\alpha}_y^*(\theta_{t,2},\phi_{t,2}) \otimes \boldsymbol{\alpha}_z^*(\phi_{t,2}) \otimes (\boldsymbol{\Phi}_{10}\boldsymbol{\alpha}_x(\theta_{r,4},\phi_{r,4}) \otimes \boldsymbol{\alpha}_y(\theta_{r,4},\phi_{r,4})+\boldsymbol{\alpha}_x(\theta_{r,4},\phi_{r,4}) \otimes \boldsymbol{\Phi}_{11}\boldsymbol{\alpha}_y(\theta_{r,4},\phi_{r,4}))).  \label{eq:partial_derivative_end}
\end{align}\normalsize
\end{figure*}
where $\boldsymbol{\Phi}_1 =-j \pi  \cos(\theta_{t,2}) \sin(\phi_{t,2}) \mathrm{diag}([-\frac{M_{\text{B},y} -1}{2}, \cdots, \frac{M_{\text{B},y} -1}{2}])$, $\boldsymbol{\Phi}_2 =-j \pi  \sin(\theta_{t,2}) \cos(\phi_{t,2}) \mathrm{diag}([-\frac{M_{\text{B},y} -1}{2}, \cdots, \frac{M_{\text{B},y} -1}{2}])$, $\boldsymbol{\Phi}_3 = j \pi \sin(\phi_{t,2})\mathrm{diag}([-\frac{M_{\text{B},z} -1}{2}, \cdots, \frac{M_{\text{B},z} -1}{2}])$, $\boldsymbol{\Phi}_4 =j \pi  \sin(\theta_{t,3}) \sin(\phi_{t,3}) \mathrm{diag}([-\frac{M_{\text{R},x} -1}{2}, \cdots, \frac{M_{\text{R},x} -1}{2}])$, $\boldsymbol{\Phi}_5 =-j \pi  \cos(\theta_{t,3}) \sin(\phi_{t,3}) \mathrm{diag}([-\frac{M_{\text{R},y} -1}{2}, \cdots, \frac{M_{\text{R},y} -1}{2}])$, $\boldsymbol{\Phi}_6 =-j \pi  \cos(\theta_{t,3}) \cos(\phi_{t,3}) \mathrm{diag}([-\frac{M_{\text{R},x} -1}{2}, \cdots, \frac{M_{\text{R},x} -1}{2}])$, $\boldsymbol{\Phi}_7 =-j \pi  \sin(\theta_{t,3}) \cos(\phi_{t,3}) \mathrm{diag}([-\frac{M_{\text{R},y} -1}{2}, \cdots, \frac{M_{\text{R},y} -1}{2}])$, $\boldsymbol{\Phi}_8 =-j \pi  \sin(\theta_{r,4}) \sin(\phi_{r,4}) \mathrm{diag}([-\frac{M_{\text{U},x} -1}{2}, \cdots, \frac{M_{\text{U},x} -1}{2}])$, $\boldsymbol{\Phi}_9 =j \pi  \cos(\theta_{r,4}) \sin(\phi_{r,4}) \mathrm{diag}([-\frac{M_{\text{U},y} -1}{2}, \cdots, \frac{M_{\text{U},y} -1}{2}])$, $\boldsymbol{\Phi}_{10} =j \pi  \cos(\theta_{r,4}) \cos(\phi_{r,4}) \mathrm{diag}([-\frac{M_{\text{U},x} -1}{2}, \cdots, \frac{M_{\text{U},x} -1}{2}])$, $\boldsymbol{\Phi}_{11} =j \pi \sin(\theta_{r,4}) \cos(\phi_{r,4}) \mathrm{diag}([-\frac{M_{\text{U},y} -1}{2}, \cdots, \frac{M_{\text{U},y} -1}{2}])$.

In this work, we only consider the angular parameter estimates for passive 3D drone localization. The transform matrix, a.k.a. Jacobian matrix, is calculated based on the geometrical relationship among the network nodes~\cite{Jiguang2023}. In its first six columns, the transform matrix $\bT\in\mathbb{R}^{3\times 10}$ has only non-zero elements in the following places: 
\begin{align}\label{bT_i1}
    &[\bT]_{15} = \partial \theta_{t,2} / \partial  x_{\text{D}}  =- \frac{\sin(\theta_{t,2})}{d_{2} \cos(\phi_{t,2})}, \\
    &[\bT]_{25} = \partial \theta_{t,2}/ \partial  y_{\text{D}}  = \frac{\cos(\theta_{t,2})}{d_{2} \cos(\phi_{t,2})}, \\
       & [\bT]_{16} = \partial \phi_{t,2} / \partial  x_{\text{D}}  =- \frac{\cos(\theta_{t,2})\sin(\phi_{t,2})}{d_{2} }, \\
     &[\bT]_{26} =\partial \phi_{t,2}/ \partial  y_{\text{D}}  =- \frac{\sin(\theta_{t,2})\sin(\phi_{t,2})}{d_{2} }, \\
     &[\bT]_{36} =\partial \phi_{t,2} / \partial  z_{\text{D}}  = \frac{\cos(\phi_{t,2})}{d_{2}},\label{bT_i6}
\end{align}
where $(x_{\text{D}},y_{\text{D}},z_{\text{D}})$ is the 3D coordinate of the drone. Similarly, we can calculate the non-zero entries within $[\bT]_{:,7:10}$. The localization error bound in terms of root mean square error (RMSE) can be characterized by 
\begin{align}\label{PEB}
 \sigma_{\hat{\bp}_\text{D}} &=  \sqrt{\mathbb{E}\{(\bp_\text{D}  -\hat{\bp}_\text{D})^\mathsf{T} (\bp_\text{D} -\hat{\bp}_\text{D})\}}  \nonumber\\
   &\geq \sqrt{\mathrm{Tr}\Big\{\big(\bT  \bJ \bT^\mathsf{T}\big)^{-1}\Big\}},
\end{align}
where $\bp_\text{D}  =  (x_{\text{D}},y_{\text{D}},z_{\text{D}})$ and $\hat{\bp}_\text{D}$ is its estimate.

\section{Drone Localization Algorithm}
For the sake of tractability, we develop a multi-stage localization approach: i) estimation of $\{\bh_4, \bh_2, \bh_3\}$, ii) extraction of $\{\theta_{t,2}, \phi_{t,2}, \theta_{t,3}, \phi_{t,3}, \theta_{r,4}, \phi_{r,4}\}$, and iii) mapping the estimates from step ii) to the 3D drone location. 




\subsection{CGD Based Channel Estimation}
In the first stage, we resort to CGD for estimating $\bh_4$, $\bh_2$, and $\bh_3$ in a sequential manner. 
The objective is to minimize 
\begin{equation}
     \min f(\bh_2, \bh_3, \bh_4) = \Big\|\bY -\sqrt{P/2} \tilde{\bH} \bar{\boldsymbol{\Omega}} - \sqrt{P/2}\hat{\bH} \bar{\bF} \Big\|_\mathsf{F}^2.
\end{equation}
The CGD approach provides the solutions for the three variables iteratively until convergence or reaching a certain stopping criterion. 
The major updating steps are summarized in the following:
\begin{align}
\bh_4^{(k)} &= \bh_4^{(k-1)} - \eta \frac{\partial f\Big( \bh_2^{(k-1)}, \bh_3^{(k-1)}, \bh_4 \Big)}{\partial \bh_4 } \bigg|_{\bh_4 = \bh_4^{(k-1)}}, \\
\bh_2^{(k)} &= \bh_2^{(k-1)} - \eta \frac{\partial f\Big( \bh_2, \bh_3^{(k-1)}, \bh_4^{(k)} \Big)}{\partial \bh_2 } \bigg|_{\bh_2 = \bh_2^{(k-1)}}, \\
\bh_3^{(k)} &= \bh_3^{(k-1)} - \eta \frac{\partial f\Big( \bh_2^{(k)}, \bh_3, \bh_4^{(k)} \Big)}{\partial \bh_3 } \bigg|_{\bh_3 = \bh_3^{(k-1)}},
\end{align}
where $\eta$ is the step size and superscript $(k)$ specifies the iteration index. The above gradient terms are further detailed below:
\begin{align}
    \frac{\partial f\Big( \bh_2^{(k-1)}, \bh_3^{(k-1)}, \bh_4 \Big)}{\partial \bh_4 } &= \bD^\mathsf{H}\bD \bh_4 - \bD^\mathsf{H} \mathrm{vec}(\bY),\nonumber\\
    \frac{\partial f\Big( \bh_2, \bh_3^{(k-1)}, \bh_4^{(k)} \Big)}{\partial \bh_2 }& = \big(\bB^\mathsf{H}\bB \bh_2 - \bB^\mathsf{H} \mathrm{vec}(\bY_1)\big)^\mathsf{T},\nonumber\\
    \frac{\partial f\Big( \bh_2^{(k)}, \bh_3, \bh_4^{(k)} \Big)}{\partial \bh_3 } & =  \big(\bC^\mathsf{H}\bC \bh_3 - \bC^\mathsf{H} \mathrm{vec}(\bY_2)\big)^\mathsf{T},\nonumber
\end{align}
where 
\begin{align}
 &\bY_1 = \bY - \sqrt{P/2} \bh_4^{(k)} \bh_3^{(k-1)}\bar{\boldsymbol{\Omega}}, \\
 &\bY_2  = \bY - \sqrt{P/2} \bh_4^{(k)} \bh_2^{(k)} \bar{\bF},\\
&\bB \;\, = \sqrt{P/2}\Big( \bar{\bF}^\mathsf{T} \otimes \bh_4^{(k)}\Big), \bC \;\, = \sqrt{P/2}\Big( \bar{\boldsymbol{\Omega}}^\mathsf{T} \otimes \bh_4^{(k)}\Big),\\
&\bD \;\,= \Big(\sqrt{P/2}\bh_3^{(k-1)}  \bar{\boldsymbol{\Omega}} + \sqrt{P/2}\bh_2^{(k-1)} \bar{\bF}\Big)^\mathsf{T} \otimes \bI_{M_\text{U}}.
\end{align}

\textbf{Remark 1}: Here, we integrate $\zeta$ into $\bh_4$ for the sake of simplicity. Intuitively, the scaling coefficient $\zeta$ will bring a significant impact on the estimation results, verified in Section~\ref{sec_Numerical_study}. 

\textbf{Remark 2}: We employ the iterative CGD method for the channel vector estimation due to its low computational complexity. Alternatively,  one can also exploit nuclear norm minimization, posing a regularized optimization problem~\cite{Shuhang2014}. However, this approach entails inevitably higher complexity, particularly when the numbers of RIS elements and BS antennas are large, leading to larger-sized matrices $\tilde{\bH}$ and $\hat{\bH}$.

\subsection{Angular Parameter Estimation via 2D Search} \label{sec_ang_par_est}
In the second stage, we extract the angular parameters from the estimate of $\bh_i$, denoted as $\hat{\bh}_i$, for $i = 2,3,4$. Based on two-dimensional (2D) search, we can find the angular parameter estimates, $\hat{\theta}_{t,i}, \hat{\phi}_{t,i}$, for $i = 2,3,4$, as follows: 
\small
\begin{align}
        \hat{\theta}_{t,2}, \hat{\phi}_{t,2} &= \argmax_{ \theta_{t,2}, \phi_{t,2} } \Big|\hat{\bh}_2\big(\boldsymbol{\alpha}_y(\theta_{t,2},\phi_{t,2}) \otimes \boldsymbol{\alpha}_z(\phi_{t,2})\big)\Big|,\nonumber\\
        \hat{\theta}_{t,3}, \hat{\phi}_{t,3} &= \argmax_{ \theta_{t,3}, \phi_{t,3} } \Big|\hat{\bh}_3\big(\boldsymbol{\alpha}_x(\theta_{t,3},\phi_{t,3}) \otimes \boldsymbol{\alpha}_y(\theta_{t,3},\phi_{t,3})\big)\Big|,   \nonumber\\
       \hat{\theta}_{r,4}, \hat{\phi}_{r,4} &= \argmax_{ \theta_{r,4}, \phi_{r,4} } \Big|\hat{\bh}_4^\mathsf{H}\big(\boldsymbol{\alpha}_x(\theta_{r,4},\phi_{r,4}) \otimes \boldsymbol{\alpha}_y(\theta_{r,4},\phi_{r,4})\big)\Big|. \nonumber 
\end{align}\normalsize

\subsection{Drone Location Mapping}
Based on the estimated angular parameters $\{\hat{\theta}_{t,2}, \hat{\phi}_{t,2}, \hat{\theta}_{t,3}, \hat{\phi}_{t,3}, \hat{\theta}_{r,4}, \hat{\phi}_{r,4}\}$, we in the last stage calculate the 3D drone location by taking into consideration the geometrical relationship between the drone and the other network nodes in the spherical coordinate system. We apply the least squares (LS) principle for mapping those angular estimates to the 3D position estimate of the drone, $\hat{\bp}_{\text{D}}$, as \cite[Section IV.D]{Jiguang2023}
\begin{equation}\label{LS_Loc}
    \hat{\bp}_{\text{D}} = \left(\bB_\text{B} + \bB_\text{R} + \bB_\text{U}\right)^{-1} \left( \bB_\text{B} \bp_{\text{B}}+\bB_\text{R} \bp_{\text{R}}+\bB_\text{U} \bp_{\text{U}}  \right),
\end{equation}
where $\bB_\text{B} \triangleq \bI_3 - \hat{\boldsymbol{\xi}}_\text{B} \hat{\boldsymbol{\xi}}_\text{B}^\mathsf{T}$ with $\hat{\boldsymbol{\xi}}_\text{B} \triangleq [\cos(\hat{\theta}_{t,2}) \cos(\hat{\phi}_{t,2}), \sin(\hat{\theta}_{t,2}) \cos(\hat{\phi}_{t,2}), \sin(\hat{\phi}_{t,2})]^\mathsf{T}$, $\bB_\text{R} \triangleq \bI_3 - \hat{\boldsymbol{\xi}}_\text{R} \hat{\boldsymbol{\xi}}_\text{R}^\mathsf{T}$ with $\hat{\boldsymbol{\xi}}_\text{R} \triangleq [\cos(\hat{\theta}_{t,3}) \cos(\hat{\phi}_{t,3})  , \sin(\hat{\theta}_{t,3}) \cos(\hat{\phi}_{t,3}), \sin(\hat{\phi}_{t,3})   ]^\mathsf{T}$, $\bB_\text{U} \triangleq \bI_3 - \hat{\boldsymbol{\xi}}_\text{U} \hat{\boldsymbol{\xi}}_\text{U}^\mathsf{T}$ with $\hat{\boldsymbol{\xi}}_\text{U} \triangleq [\cos(\hat{\theta}_{r,4}) \cos(\hat{\phi}_{r,4})  , \sin(\hat{\theta}_{r,4}) \cos(\hat{\phi}_{r,4}), \sin(\hat{\phi}_{r,4})   ]^\mathsf{T}$, and $ \bp_\text{B}$, $\bp_\text{R}$, and $\bp_\text{U}$ are the 3D coordinates of the BS, the RIS, and UE, respectively.

\section{Simulation  Results}\label{sec_Numerical_study}
\begin{figure}[t]
	\centering
\includegraphics[width=0.82\linewidth]{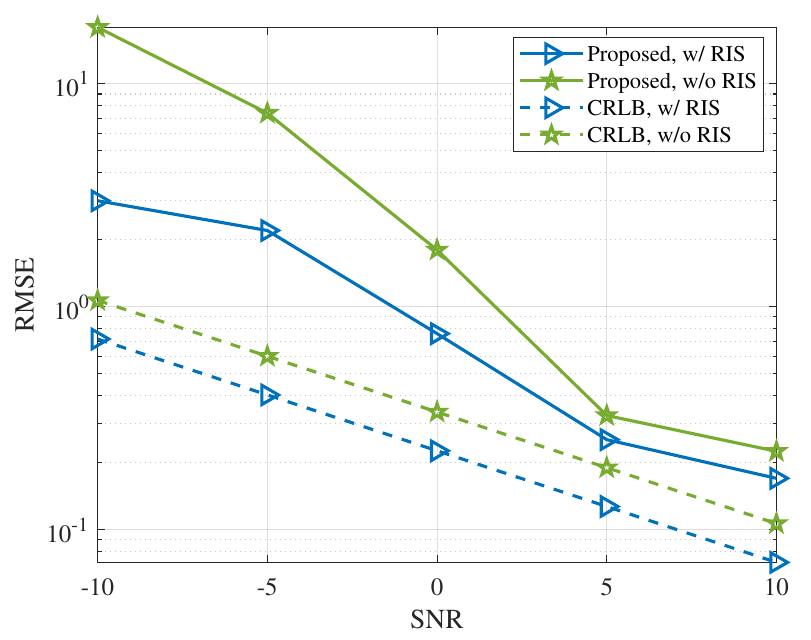}
	\caption{RMSE performance with and without RIS, where $K = 60$ and $\zeta = 1$.}
 \label{Training_overhead60_with_without_RIS}
		 \vspace{-0.5cm}
\end{figure}

In this section, we evaluate the performance of the RIS-assisted drone localization systems using the following parameter configuration: $M_\text{B} = 8\times 8, M_\text{U} = 4 \times 4$, and $M_\text{R}= 6\times 6$. The locations of the BS, RIS, drone, and UE are specified as $(0,0,26)$, $(0,0.5,25.5)$, $(3,3,30)$, and $(2,2, 24)$, respectively. The path loss $\rho_i$ is defined as $ \rho_i = d_i^{\Gamma}$ with $\Gamma =2$. Leveraging geometric relationships, we compute angular parameters for each of the five individual channels. The channel bandwidth $B$ is fixed as $20$ MHz, yielding a noise variance of $\sigma^2 = -174\; \text{dBm} + 10 \log_{10}(B) = -101$ dBm. The signal-to-noise ratio (SNR) $P/\sigma^2$ is varied from $-10$ dB to $10$ dB. A straightforward benchmark involves comparing drone localization performance without an RIS, relying solely on two reference nodes within the system. The simulation results, depicted in Fig.~\ref{Training_overhead60_with_without_RIS}, reveal a significant improvement with the introduction of an RIS, particularly evident in the low SNR regime. In the legend, ``Proposed'' stands for the developed practical localization algorithm based on CGD.  




\begin{figure}[t]
	\centering
\includegraphics[width=0.82\linewidth]{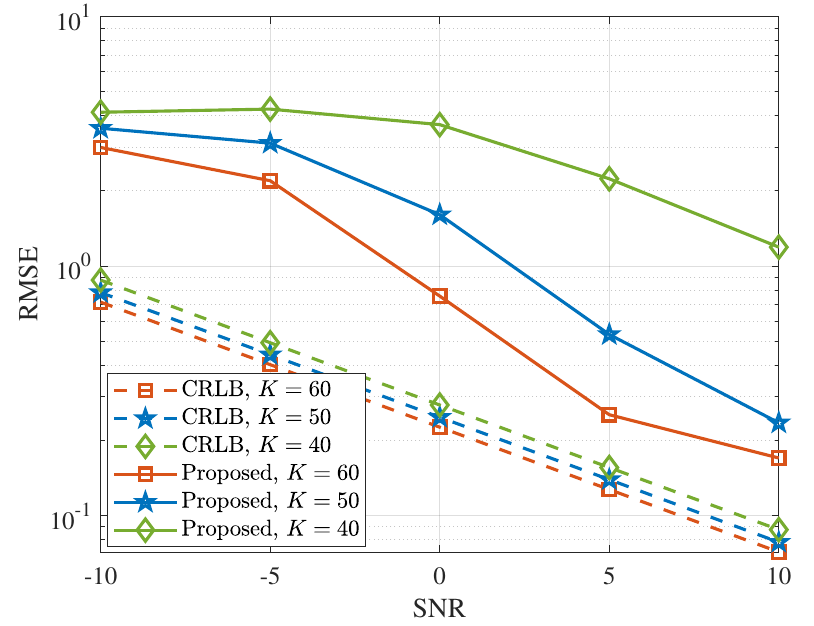}
	\caption{RMSE performance with various $K$ values and $\zeta = 1$.}
 \label{Drone_Localization_Training_Overhead}
		\vspace{-0.5cm}
\end{figure}

We examine the impact of training overhead on performance. Various levels of training overhead, i.e., $K$ values, are considered. The simulation results are illustrated in Fig.~\ref{Drone_Localization_Training_Overhead}. As expected, higher training overhead leads to improved drone localization performance. 

\begin{figure}[t]
	\centering
\includegraphics[width=0.82\linewidth]{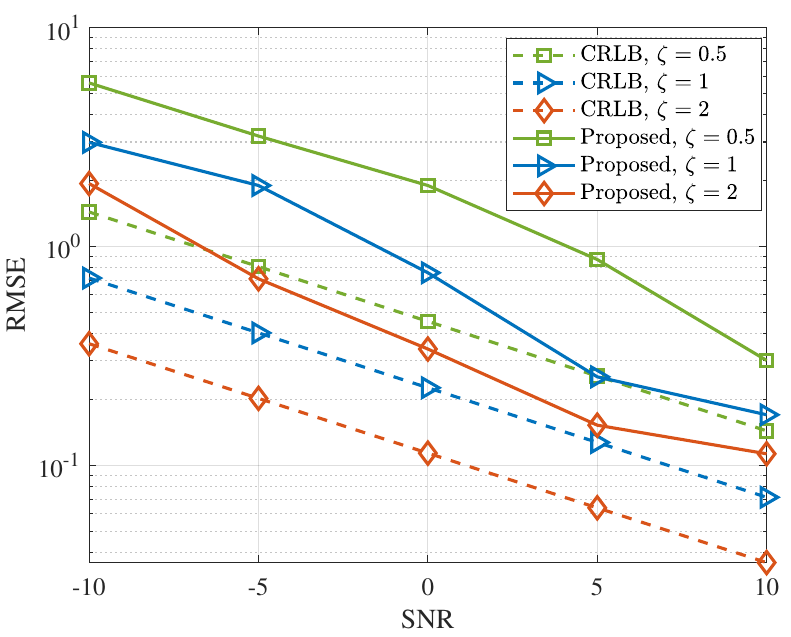}
	\caption{RMSE performance with $K = 60$ and $\zeta \in \{0.5, 1, 2\}$.}
		\label{RMSE_RCS}
		 \vspace{-0.5cm}
\end{figure}

The RCS values play a pivotal role in affecting the performance. Hence, we evaluate the following three cases: $\zeta \in\{ 0.5,1,2\}$. The simulation results are depicted in Fig.~\ref{RMSE_RCS}. Notably, RCS values exhibit a substantial impact on the accuracy of angular parameter estimation (as discussed in Section~\ref{sec_ang_par_est}), which in turn affects the passive 3D drone localization performance. 

\section{Conclusions}
In this paper, we have investigated device-free drone localization within RIS-assisted mmWave MIMO networks, specifically focusing on scenarios where direct communication between the drone and the BS is absent. We have theoretically analyzed the performance limits via Fisher information analysis as well as CRLB and developed a practical localization algorithm based on CGD. Furthermore, we have examined various system parameters to offer insights for the prospective deployment of such systems. The optimal power allocation among the BS beams will be considered in our future work. 

\section*{Acknowledgement}
Aymen Fakhreddine's contribution is funded by the Austrian Science Fund (FWF\,--\,Der Wissenschaftsfonds) under grant ESPRIT-54 (Grant DOI: 10.55776/ESP54).

\bibliographystyle{IEEEtran}
\bibliography{IEEEabrv,Ref}

\end{document}